\documentclass[12pt]{iopart}

\usepackage{subfigure}
\usepackage{graphicx}
\usepackage{color}
\begin{document}

\title[Eigenstate thermalization hypothesis, time operator, \\ and extremely quick relaxation of fidelity]{Eigenstate thermalization hypothesis, time operator, \\ and extremely quick relaxation of fidelity}

\author{Takaaki Monnai}

\address{Department of Materials and Life Sciences, Seikei University, Tokyo, 180-8633, Japan}
\ead{monnai@st.seikei.ac.jp}
\vspace{10pt}
\begin{indented}
\item[]February 2018
\end{indented}

\begin{abstract}
The eigenstate thermalization hypothesis (ETH) insists that for nonintegrable systems each energy eigenstate accurately gives microcanonical expectation values for a class of observables. As a mechanism for ETH to hold,
we show that the energy eigenstates are superposition of uncountably many quasi eigenstates of operationally defined ``time operator", which are thermal for thermodynamic isolated quantum many-body systems and approximately orthogonal in terms of extremely short relaxation time of the fidelity. In this way, our scenario   provides a theoretical explanation of ETH.  
\end{abstract}

%
%
%
%
%

\section{Introduction}
Recently, considerable attention has been paid to the foundation of statistical mechanics on the basis of intrinsic thermal nature of individual pure states.
The long standing fundamental problems are to derive the principle of equal weight and to explain the mechanism of irreversible thermalization in terms of isolated quantum many-body systems\cite{Rigol1,Vidmar1,D'Alessio1,Gogolin1}.      

In particular, typicality shows that a pure state uniform randomly sampled with respect to the Haar measure from an appropriate energy shell well represents the microcanonical ensemble\cite{Sugita1,Reimann1,Lebowitz1,Popescu1}, which provides a simple scenario to justify the principle of equal weight: Fix a set of observables, then the majority of the pure states in the Hilbert space are similar each other when calculating the expectation values. 
Thus, we may superpose them with an almost arbitrary weight, which includes the case of equal weight. 

Regarding the thermalization, several different approaches have been studied such as the restriction to the macroscopic observables\cite{vonNeumann1,Goldstein2}, the general evaluation of relaxation time\cite{Monnai1,Goldstein1,Reimann2,Santos1}, the Eigenstate thermalization hypothesis (ETH)\cite{Rigol1,Parma1,Deutsch1,
Reimann3,Anza1,Anza2}, and dynamical experiments in autonomous cold atomic systems\cite{Kinoshita1,Schmiedmayer1,Gring1}. Among these various issues, we focus on the foundation of ETH, which insists that each energy eigenstate well represents the microcanonical ensemble for nonintegrable systems, i.e. its expectation values of a class of observables well agree with the microcanonical averages. By requiring this property and non-degeneracy condition, arbitrary initial pure states equilibrate for the long time average of expectation values of fixed observables.
ETH has been discussed in terms of the nonintegrability\cite{Rigol1,Alba1,Shankar1}, partly because the relaxation property is considered to be sensitive to the presence of integrals of motion.
However, Refs. \cite{Alba1,Shankar1,Vidmar2} indicate that most energy eigenstates of integrable systems are thermal. Such an intrinsic thermal nature shared by most  energy eigenstates of integrable systems is often called weak ETH. By considering the observables of  a small subsystem, the ETH resembles typicality, though there is still a possibility that the deviations from microcanonical ensemble average of typical states and energy eigenstates are quantitatively different. We will numerically evaluate the deviations later. 
 
Let us try to understand the mechanism of ETH in terms of typicality. Our starting point is  
to seek a relevant basis $\{|\phi_n\rangle\}$, which is thermal and each energy eigenstate is a superposition of sufficient number of orthonormal states, 
\begin{equation}
|E_n\rangle=\sum_{m=1}^d c_m|\phi_m\rangle. \label{superposition1}
\end{equation}
Here, $d={\rm dim}{\cal H}_{[E,E+\Delta E]}$ is the dimension of the energy shell, and $c_n={\cal O}(\frac{1}{\sqrt{d}})$.

In this paper, we show a scenario for ETH to hold by explaining that the quasi eigenstates $|\Psi(t)\rangle$ of ``time operator" $\hat{T}$ are relevant basis by considering a thermodynamic system where $|\Psi(t)\rangle$ thermalizes and stays in equilibrium. We remark that the ``time operator" is constructed in (\ref{time1}) via spectral decomposition and is approximately canonical conjugate to the Hamiltonian up to a  constant due to long time cutoff, however, proper definitions of ``time operator" is in general still controversial.    
Instead of asking which definition is the best, we explore a foundation of ETH by introducing ``time operator" as (\ref{time1}) and its quasi eigenstates, which are approximately orthogonal with the use of extremely quick relaxation of the fidelity\cite{Monnai1,Goldstein1,Reimann2,Santos1}. Note that such a quasi orthogonality is analogous to that of the coherent state, and also used in quantum non demolition measurement\cite{Milburn1,Milburn2}.  
In particular, we show that each energy eigenstate can be expressed as a superposition of mutually almost orthogonal many pure states , which are considered as thermal. We remark that Ref.\cite{Anza1} quantifies the degree of superposition with the use of Shannon entropy, which is basis dependent and maximized to guarantee ETH. Subsequently, Ref.\cite{Anza2} addresses the issue to specify a class of observables such as local and extensive quantities that satisfy ETH in terms of mutually unbiased basis with respect to Hamiltonian. Mutually unbiasedness can be  regarded as a generalization of the concept of canonical conjugate, which is significant in our argument, and thus \cite{Anza1,Anza2} are considered to be related to the present work. The main difference is that in the present article, the quasi eigenstates of ``time operator" are unbiased with respect to the Hamiltonian, however, we do not attempt to apply ETH to the ``time operator"" itself. Instead, we explain that the vast majority of the quasi eigenstates of ``time operator" are regarded as thermal with the aid of the typicality\cite{Sugita1,Reimann1,Lebowitz1,Popescu1} and the assumption of equilibration by considering observables of a subsystem. 
Then, the energy eigenstates are regarded as thermal.

This paper is organized as follows.
In Sec. 2, we express the energy eigenstates in terms of quasi eigenstates of ``time operator", and
 explore the orthogonality and thermal nature of quasi eigenstates of the ``time operator". In Sec. 3, we numerically verify the approximate orthogonality, and thermal nature of quasi eigenstates and ETH.
Sec. 4 is devoted to a summary.
\section{Time evolved states}
Suppose that the Hamiltonian $\hat{H}=\sum_{n=1}^dE_n|E_n\rangle\langle E_n|$ has a thermodynamic density of the states $\Omega(E)=e^{N\phi(\frac{E}{N})}$, i.e. the entropy is additive for large system size $N$ and $\phi(x)$ is concave. We also assume that the eigenenergies are not degenerated $E_n\neq E_m$ ($n\neq m$). These assumptions will be used to  the orthogonality and completeness of time evolved states.  
We randomly choose a state $|\Psi\rangle=\sum_{n=1}^dc_n|E_n\rangle$ from an energy shell ${\cal H}_{[E,E+\Delta E]}$, and consider the state $|\Psi(t)\rangle=e^{-i\frac{\hat{H}}{\hbar}t}|\Psi\rangle$ at time $t$. 
   
As a typical superposition, we choose $c_n=\frac{1}{\sqrt{d}}$, since the mean value of the coefficients with respect to the Haar measure is calculated by $\overline{|c_n|^2}=\frac{1}{d}$. Here, $\overline{\cdot}$ denotes the average over $c_n$.
Note that for random sampling of a state, the absolute values of the coefficients are $\frac{1}{\sqrt{d}}$ plus small fluctuation. By taking into account the small fluctuation, our main point namely to express the energy eigenstates as a superposition of sufficiently many thermal basis is unchanged, though the factor $\sqrt{d}$ should be replaced by $\frac{1}{c_n}$ in (\ref{Fourier1}) and the commutation relation (\ref{commutator1}) is slightly modified.      
 Here, we set  $c_n$ real, since the phase factor at $t=0$ can be absorbed to the definition of energy eigenstates $|E_n\rangle$.  
Then, we can show that $\langle\Psi(t)|\Psi(s)\rangle\cong 0$ for $|t-s|$ larger than the time resolution $\tau=\frac{2\pi\hbar}{\Delta E_{\rm eff}}$\cite{Monnai1}, which will be explained later.

Let us formally define the ``time operator" as
\begin{equation}
\hat{T}=\frac{d}{T}\int_0^T t|\Psi(t)\rangle\langle\Psi(t)|dt, \label{time1}
\end{equation}
where we consider a large but finite time $T$. 
It is remarked that by introducing the infinitesimally small cut off frequency $\epsilon$\cite{Thirring1,Monnai3}, we may alternatively define as
\begin{equation}
\hat{T}=\lim_{\epsilon\rightarrow 0}\epsilon\int_0^\infty te^{-\epsilon t}|\Psi(t)\rangle\langle\Psi(t)|dt. \label{time2}
\end{equation} 
Despite the uncertainty relation $\Delta E\Delta t\geq \frac{\hbar}{2}$, it is well-known that ``time operator" which is canonical conjugate to the Hamiltonian does not exist as an observable\cite{Olkhovsky1,Wang1,Pegg1,Suskind1}, partly because the Hamiltonian is bounded below. 
On the other hand, ``time operator" defined by (\ref{time1}) approximately satisfies the commutation relation up to a boundary constant just as in the case of ``phase operator"\cite{Pegg1}  
\begin{equation}
[\hat{H},\hat{T}]=-i\hbar\hat{I}_T+i\hbar d|\Psi(T)\rangle\langle\Psi(T)| \label{commutator1}
\end{equation}  
with $\lim_{T\rightarrow\infty}\hat{I}_T=\sum_{n=1}^d|E_n\rangle\langle E_n|$.

Actually, it is straightforward to formally show (\ref{commutator1}) with the use of  $\hat{H}|\Psi(t)\rangle=i\hbar\frac{\partial}{\partial t}|\Psi(t)\rangle$, and a partial integral
\begin{eqnarray}
&&[\hat{H},\hat{T}] \nonumber \\
&=&\frac{d}{T}\int_0^T t\left(i\hbar(\frac{\partial}{\partial t}|\Psi(t)\rangle)\langle\Psi(t)|+i\hbar|\Psi(t)\rangle\frac{\partial}{\partial t}(\langle\Psi(t)|)\right) \nonumber \\
&=&i\hbar\frac{d}{T}(t|\Psi(t)\rangle\langle\Psi(t)|)|_0^{T} \nonumber\\
&&-i\hbar\frac{1}{T}\int_0^Te^{-\frac{i}{\hbar}(E_n-E_m)t}\sum_{n,m}|E_n\rangle\langle E_m| \nonumber \\
&=&-i\hbar\hat{I}_T+i\hbar d|\Psi(T)\rangle\langle\Psi(T)|, \label{commutation2}
\end{eqnarray}
where $\hat{I}_T=\frac{1}{T}\int_0^Te^{-\frac{i}{\hbar}(E_n-E_m)t}\sum_{n,m=1}^d|E_n\rangle\langle E_n|$ converges to $\hat{I}=\sum_{n=1}^d|E_n\rangle\langle E_n|$ as $T\rightarrow\infty$ from the nondegeneracy assumption. From (\ref{time2}), a similar calculation formally removes the boundary term at $t=T$ with the use of the nondegeneracy, however, in this case $\hat{T}$ is not bounded. 
In our case, the diagonal basis $|\Psi(t)\rangle$ of (\ref{time1}) are eigenstates of $\hat{T}$ approximately, since the orthogonality holds up to the time resolution $\tau$, which is in marked contrast to the case of mechanical observables. We will explain that the time resolution $\tau$ is extremely short for thermodynamic systems. 

We can express the energy eigenstates by the inverse Fourier transform   
\begin{equation}
|E_n\rangle=\lim_{T\rightarrow\infty}\frac{\sqrt{d}}{T}\int_0^Te^{\frac{i}{\hbar}E_n t}|\Psi(t)\rangle dt. \label{superposition2}
\end{equation}
Eq. (\ref{superposition2}) shows that the energy eigenstates are superposition of continuously many quasi eigenstates of ``time operator", which approximately satisfies the orthogonality(\ref{orthogonal1})\cite{Monnai1}. In the next section, we also explain that the quasi eigenstate $|\Psi(t)\rangle$ typically well represents the microcanonical state, and is considered as a relevant basis to discuss the foundation of ETH.  
Next, we analytically explore the quasi orthogonality and equilibrium nature of the basis $|\Psi(t)\rangle$.  
First, we recall the calculation of the fidelity detailed in \cite{Monnai1}  (see also \cite{Reimann2,Santos1}). 
By using 
\begin{equation}
|\Psi(t)\rangle=\frac{1}{\sqrt{d}}\sum_{n=1}^de^{-\frac{i}{\hbar}E_nt}|E_n\rangle, \label{Fourier1}
\end{equation}
 the inner product of states at time $t$ and $s$ satisfies
\begin{equation}
\langle \Psi(t)|\Psi(t)\rangle=1 \label{normalization1}
\end{equation}
for $t=s$ and
\begin{eqnarray}
&&\langle \Psi(t)|\Psi(s)\rangle \nonumber \\
&=&\frac{1}{d}\sum_{n=1}^d e^{\frac{i}{\hbar}E_n(t-s)} \nonumber \\
&\cong&\frac{1}{d}\int_E^{E+\Delta E}e^{-\frac{i}{\hbar}E'(t-s)}\Omega(E')dE' \nonumber \\
&=&\frac{1}{d\Delta E}\int_0^{\Delta E}e^{-\frac{i}{\hbar}(E+\Delta E-x)(t-s)+\log(\Omega(E+\Delta E)\Delta E)-\beta x-\frac{\beta^2}{2C_V}x^2+{\cal O}(\frac{1}{N})}dx \nonumber \\
&\cong& e^{\frac{i}{\hbar}(E+\Delta E)(t-s)}\frac{1-e^{-(\beta-\frac{i}{\hbar}(t-s))\Delta E_{\rm eff}}}{\beta\tilde{\Delta}E+\frac{i}{\hbar}\tilde{\Delta}E(t-s)}, \label{orthogonal1}
\end{eqnarray}  
where the discrete sum is evaluated as integral with the use of the density of the states $\Omega(E')$, which makes the spectrum of eigenenergy continuous and the dynamics is supposed to be irreversible. At this step, the recurrence phenomena at extremely long time is omitted. Such a continuous approximation accurately holds as shown in Fig. 1.  We expand the density of the states as $\log\Omega(E+\Delta E-x)=\log\Omega(E+\Delta E)-\beta x-\frac{\beta^2}{2C_V}x^2+{\cal O}(\frac{1}{N^2})$ with the inverse temperature $\beta=\frac{d}{dE}\log\Omega(E)|_{E+\Delta E}$, the heat capacity $C_V$, and the system size $N$. Here, we set the Boltzmann constant unity, and introduced an effective energy widths $\tilde{\Delta} E=\frac{d}{\Omega(E+\Delta E)}$ and $\Delta E_{\rm eff}(\leq \Delta E)$. In particular, $\Delta E_{\rm eff}$ is chosen so that the linear approximation of $\log\Omega(E+\Delta E-x)=\log\Omega(E+\Delta E)-\beta x-\frac{\beta^2}{2C_V}x^2+{\cal O}(\frac{1}{N})$ holds in $[E+\Delta E-\Delta E_{\rm eff},E+\Delta E]$. Here, we evaluate $\Delta E_{\rm eff}$ from the condition that the absolute value of the first order term $\beta x$ is much larger than that of the second order $\frac{\beta^2}{2C_V}x^2$ for $x=\Delta E_{\rm eff}$, which yields $C_V\gg\beta \Delta E_{\rm eff}$.     For thermodynamic density of the states, the energy width $\Delta E_{\rm eff}$ is considered to be the same order as $\frac{1}{\beta}$.
Since the heat capacity is proportional to the system size, we can accurately calculate the integral up to the first order of $x$.
At time $\tau=\frac{2\pi\hbar}{\Delta E_{\rm eff}}$, the inner product (\ref{orthogonal1}) becomes considerably small\cite{Monnai1}, which is ${\cal O}(\frac{1}{\sqrt{d}})$ and the states $|\Psi(t)\rangle$ and $|\Psi(s)\rangle$ are almost orthogonal for $|t-s|\geq\tau$. 
We also remark that from the unitarity, the short time expansion of the fidelity $|\langle\Psi|e^{-\frac{i}{\hbar}\hat{H}t}|\Psi\rangle|^2=1-{\rm Var}[\hat{H}]t^2+{\cal O}(t^4)$ suggests that the decay rate of the fidelity is determined by the energy fluctuation ${\rm Var}[\hat{H}]=\langle\Psi|(\hat{H}-\langle\Psi|\hat{H}|\Psi\rangle)^2|\Psi\rangle$, which is compatible to our evaluation of $\tau$. It is also well-known that for long time regime, the fidelity shows power-law decay by Pailey-Wiener theorem for Fourier-Laplace transformation.  Meanwhile, the exponential decay is observed for the time scale of our interest. 
 
Here, we discuss some property of the ``time operator".  
The operator  
\begin{equation}
|\Psi(t)\rangle\langle \Psi(t)|=\frac{1}{d}\sum_{n=1}^d\sum_{l=1}^de^{-i(E_n-E_l)t}|E_n\rangle\langle E_l| \label{Fourier2}
\end{equation}
can be regarded as a projection onto $|\Psi(t)\rangle$, i.e. $|\Psi(t)\rangle\langle\Psi(t)|\Psi(s)\rangle$ is proportional to $|\Psi(t)\rangle$, and is nonnegligible for $|t-s|\leq\tau$ from the quasi orthogonality. 
Note that the projection operator (\ref{Fourier2}) can be used for measurement of time as approximately projection to $|\Psi(t)\rangle$:
Given a state $|\Psi(t)\rangle$ with unknown $t$, such projection determines $t$ with an accuracy $\tau$.
Repeating this thought experiment many times with randomly distributed $t$, we actually obtain the spectral fluctuation.
The projection operator (\ref{Fourier2}) also satisfies the completeness
\begin{equation}
\lim_{T\rightarrow}\frac{d}{T}\int_0^T|\Psi(t)\rangle\langle\Psi(t)|dt=\sum_{n=1}^d|E_n\rangle\langle E_n|. \label{integral1}
\end{equation}

\section{Numerical simulation}
Let us explore the quasi orthogonality for quasi eigenstates $|\Psi(t)\rangle$. Regarding the quasi orthogonality, more detailed calculation is shown in \cite{Monnai1}. 
For concreteness, we first consider one-dimensional Ising model in a magnetic field\cite{Shankar1,Monnai2} ${\bf B}=(\alpha,0,\gamma_i)$ where $\gamma_i$ is the $z$-component at $i$-th site. 
The Hamiltonian for $N$ site is 
\begin{equation}
\hat{H}=-J\sum_{j=1}^{N-1}\hat{\sigma}_j^z\hat{\sigma}_{j+1}^z+\alpha\sum_{j=1}^N\hat{\sigma}_j^x+\sum_{j=1}^N\gamma_j\hat{\sigma}_j^z. \label{spin1}
\end{equation}
For $N=9$ and $N=10$, we choose an energy shell ${\cal H}_{[E,E+\Delta E]}$ as a subspace spanned by eigenstates $|E_n\rangle$ with (a)$151\leq n\leq 200$ and (b)$251\leq n\leq 300$. We set the parameters $J=1$ and $\alpha=1$. On the other hand, we also explored various choices of $\gamma_j$ such as uniform case $\gamma_j=0.5$ ($\gamma=0$ corresponds to the integrable case), randomly distributed case $\gamma_j\in[0,\Delta]$ with $\Delta=0.5, 1$. Here, the eigenenergies are in increasing order $E_1\leq E_2\leq E_3\leq...$. 
For the case of $\gamma_j=0.5$, the inverse temperature, energy width, and the effective energy fluctuation are (a)$\beta=0.2$, $\Delta E=0.908814$, $\tilde{\Delta}E=0.85233$ and (b)$\beta=0.12$, $\Delta E=0.722721$, $\tilde{\Delta}E=0.683753$. In these cases, we can take $\Delta E_{\rm eff}=\Delta E$, since the linearization of the entropy $\log\Omega(E+\Delta E-x)$ holds for the entire shell. 

We illustrate the time evolution of the fidelity $F(t)=|\langle\Psi|\Psi(t)\rangle|^2$ for the  case (a) in Fig. 1. The result for the case (b) is similar. We compare the exact $F(t)$ (blue curve) and approximation $\frac{\beta^2}{1+e^{-2\beta\Delta E_{\rm eff}}-2e^{-\beta\Delta E_{\rm eff}}}\frac{1+e^{-2\beta\Delta E_{\rm eff}}-2e^{-\beta\Delta E_{\rm eff}}\cos\Delta E_{\rm eff}t}{\beta^2+t^2}$ (red broken line) calculated from Eq. (\ref{orthogonal1}), where we set $\hbar=1$.
Note that the relaxation time $\tau$ is quite general\cite{Monnai1} and is the same order as the Boltzmann time $2\pi\beta\hbar$ for macroscopic systems\cite{Goldstein1}, which is extremely short at room temperature $\tau\sim 10^{-12}s$. Therefore, we can regard the basis $|\Psi(t)\rangle$ $(t\geq 0)$ in the expansion (\ref{superposition2}) are mutually orthogonal. 
\begin{figure}
\center{
\includegraphics[scale=0.6]{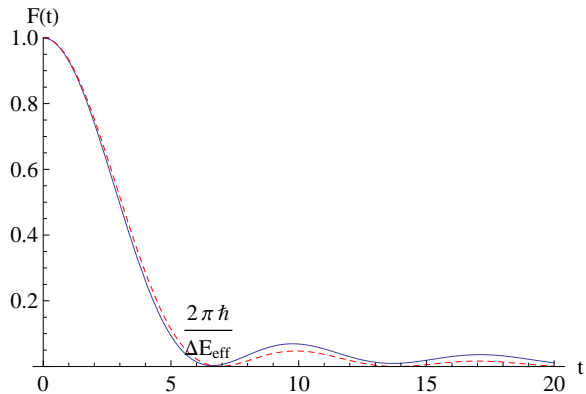}
}
\caption{The time evolution of the fidelity $F(t)$ (blue curve)for $151\leq n\leq 200$, $J=\alpha=1$, $\gamma_j=0.5$. The red broken line shows the approximation obtained by (\ref{orthogonal1}).} \end{figure}  

Next, we verify the thermalization of the quasi eigenstates of $\hat{T}$, i.e. the basis $|\Psi(t)\rangle$ well represent the microcanonical state for most $t\in[0,\infty]$. For this purpose, it is necessary to calculate the  expectation values of a class of observables $\hat{A}$ for $|\Psi(t)\rangle$ and compare with those of the microcanonical ensemble.
Theoretically, $|\Psi(t)\rangle$ describes thermal equilibrium for most $t$ according to the typicality\cite{Sugita1,Reimann1} and the unitarity of time evolution.
Numerically, we investigate the expectation values of arbitrary observables defined on the left-most $m$ sites $\hat{A}_m$ \cite{Lebowitz1,Popescu1,Parma1,Alba1}. Thus, we calculate the Hilbert-Schmidt distance $\Delta\hat{\rho}(t)=\|\hat{\rho}_m(t)-\hat{\rho}_0\|$ between the reduced density matrices $\hat{\rho}_m(t)={\rm Tr}_{N-m}|\Psi(t)\rangle\langle\Psi(t)|$ and $\hat{\rho}_0={\rm Tr}_{N-m}\frac{1}{d}\sum_{n=1}^d|E_n\rangle\langle E_n|$. Here, ${\rm Tr}_{N-m}$ stands for the partial trace for the right-most $N-m$ sites. 

In Fig. 2(a), we show the time dependence of the deviation from equilibrium $\Delta\hat{\rho}(t)$ for $1\leq m\leq 3$ (inset illustrates the time average of the deviation for $1\leq m\leq 5$) and the energy shell $[E_{151}, E_{200}]$. We note that the deviation $\Delta\hat{\rho}(t)$ is roughly upper bounded by $\sqrt{\frac{d_s}{d}}$ with $d_s=2^{m}$. Indeed, we can calculate the variance of ${\rm Tr}_{N-m}|\Phi\rangle\langle\Phi|$ as 
\begin{eqnarray}
&&\overline{({\rm Tr}_{N-m}|\Phi\rangle\langle\Phi|-{\rm Tr}_{N-m}\overline{|\Phi\rangle\langle\Phi|})^2} \nonumber \\
&=&\frac{1}{d(d+1)}(\sum_{i=1}^{d_s}\sum_{i'=1}^{d_s}\sum_{j\in\Delta_i\cap \Delta_{i'}}1-\frac{1}{d}\sum_{i=1}^{d_s}\sum_{j\in\Delta_i}\sum_{j'\in\Delta_{i'}}1) \label{variance1}
\end{eqnarray}
for a typical state $|\Phi\rangle=\sum_{i=1}^{d_s}\sum_{j\in\Delta_i}c_{ij}|E_i^{(s)}\rangle|E_j^{(r)}\rangle$ under the assumption of weak coupling.
Here, $\overline{\cdot}$ stands for the uniform average over coefficients $c_{ij}$\cite{Monnai3}, $E_i^{(s)}$ and $E_j^{(r)}$ are the local eigenenergy of the left most $m$ sites and the right most $N-m$ sites, and $\Delta_i=\{j|E-E_i^{(s)}\leq E_j^{(r)}\leq E-E_i^{(s)}+\Delta E\}$ denotes the set of excitation numbers of the right-most $N-m$ sites. By upper-bounding $\sum_{j\in\Delta_i\cap \Delta_{i'}}\leq d_r^2$ 
with the dimension of the right-most $N-m$ sites $d_r$, we can evaluate the variance as smaller than $\frac{d_s}{d}$.

Aside from the Hilbert-Schmidt distance from equilibrium, we numerically calculated the subsystem size $m$ dependences of bipartite entanglement entropies\cite{Vidmar2} of the superposition $|\Psi(t)\rangle$ and energy eigenstates $|E_n\rangle$. In Fig 3, we show the von Neumann entropy $S$ of the reduced state ${\rm Tr}_{N-m}|\Psi(t)\rangle\langle\Psi(t)|$ at $t=0,1,2,..,99$ and ${\rm Tr}_{N-m}|E_n\rangle\langle E_n|$ for $151\leq n\leq 200$ with and without disorder.

Then, we also calculated the distance between $\hat{\tau}_m^{(n)}={\rm Tr}_{N-m}|E_n\rangle\langle E_n|$ and the averaged state in Fig. 2(b). To quantitatively compare the Fig. 2(a) and (b), we define $\Delta\hat{\tau}$ and $\Delta\hat{\rho}$ as the sample average of distances $\|\hat{\tau}_m^{(n)}-\hat{\rho}_0\|$ for $151\leq n\leq 200$ and the time average of $\|\hat{\rho}_m(t)-\hat{\rho}_0\|$ for $0\leq t\leq 100$.  In Fig. 4, we show the subsystem size $m$ dependences of $\Delta\hat{\rho}$(blue curve) and $\Delta\hat{\tau}$(red curve) for the case of uniform magnetic field $\gamma_j=0.5$, and randomly sampled $\gamma_j$ from $[0,\Delta]$ with $\Delta=0.5,1$. The deviations $\Delta\hat{\rho}$ and $\Delta\hat{\tau}$ well agree for all the three cases. 
This fact means that ETH holds with the same accuracy as the thermal property of $|\Psi(t)\rangle$ for these cases.

\begin{figure}
\center{
\subfigure[]{\includegraphics[scale=0.6]{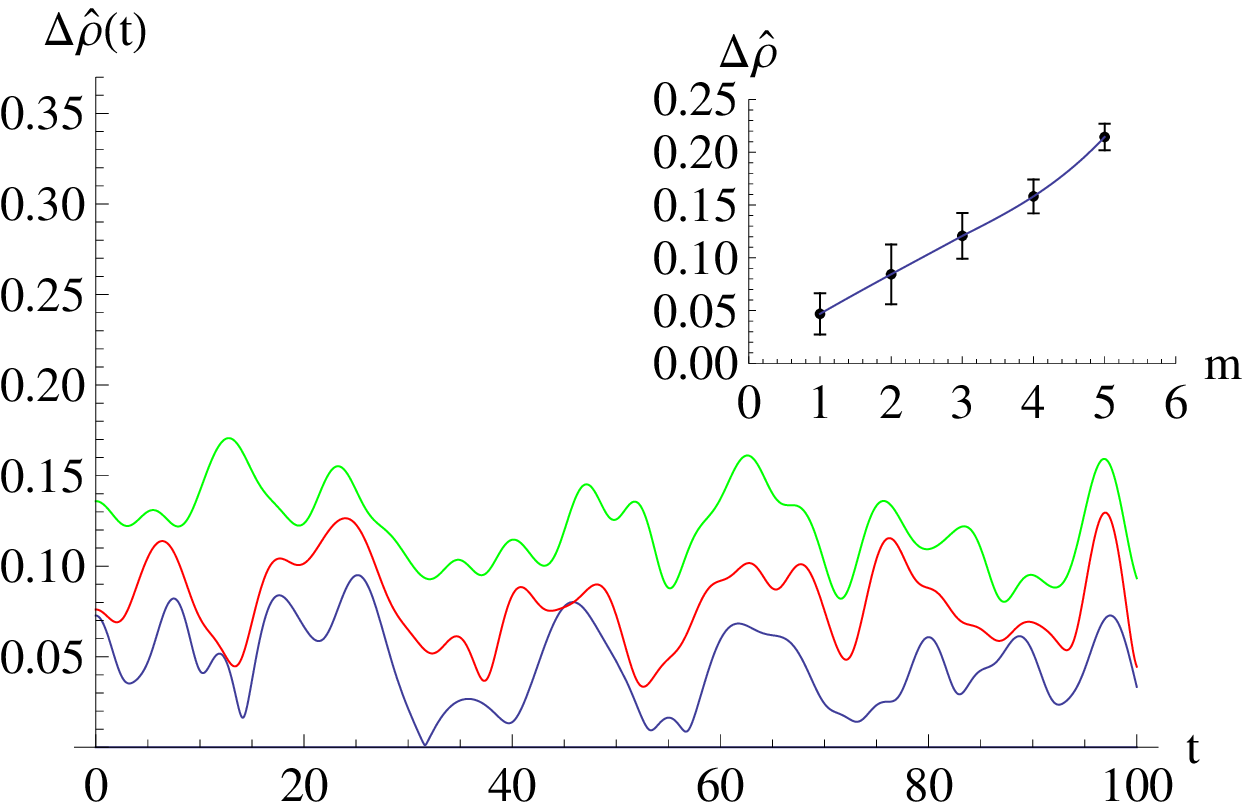}}
\subfigure[]{\includegraphics[scale=0.6]{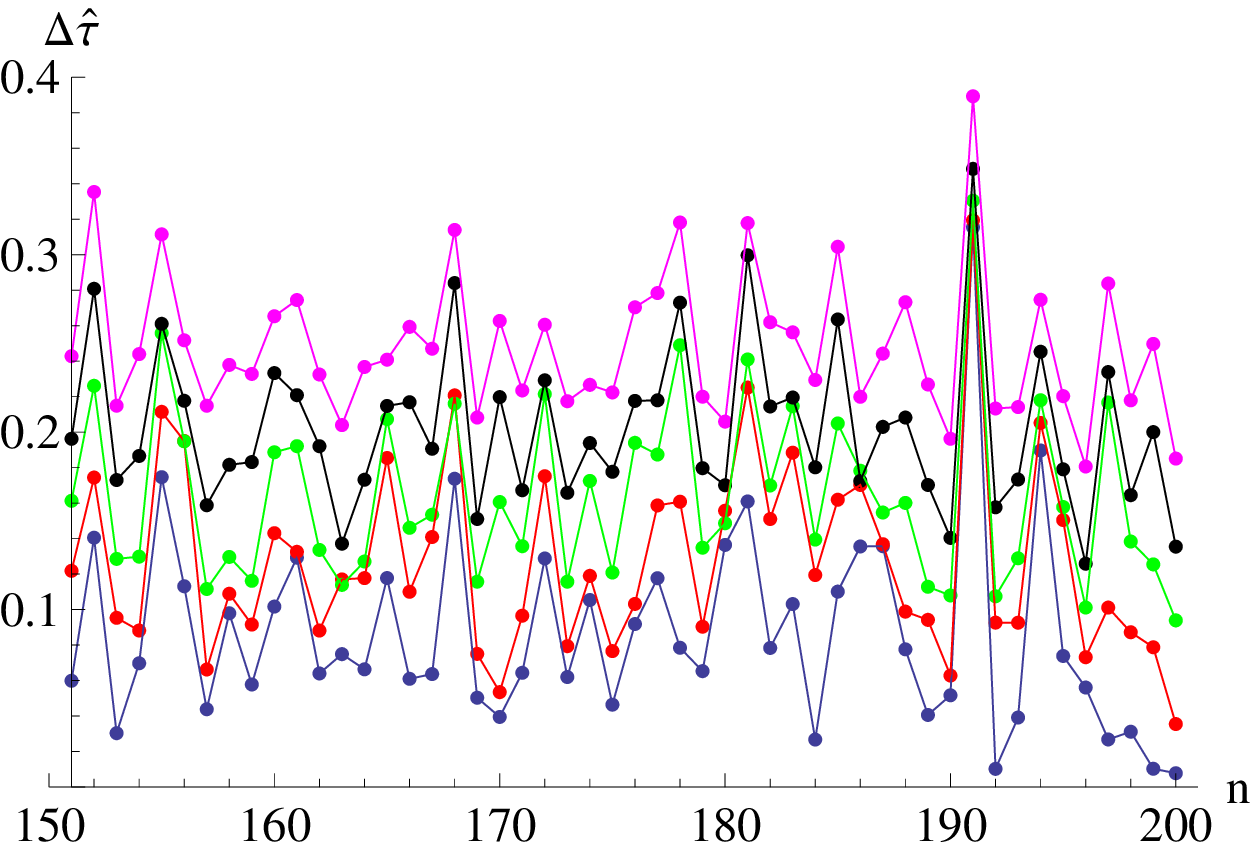}}
}
\caption{(a)The time dependence of the deviation from equilibrium for $1\leq m\leq 3$ calculated by the Hilbert-Schmidt distance between the reduced state $\hat{\rho}_m(t)$ and the averaged state $\hat{\rho}_0$. Inset shows the mean values of the distance for $1\leq m\leq 5$.
We used $|E_n\rangle$ $(151\leq 200)$, and other parameters are the same as Fig. 1. (b) The Hilbert-Schmidt distance $\Delta\hat{\tau}_n$ between the reduced energy eigenstates $\tau_n={\rm Tr}_{N-m}|E_n\rangle\langle E_n|$ and the averaged state $\hat{\rho}_0$ for $1\leq m\leq 5$. The mean square root $\Delta\hat{\tau}$ of $\Delta\hat{\tau}_n^2$ agrees with the deviation $\Delta\hat{\rho}$ of inset of Fig. 2(a) for $1\leq m\leq 5$. }
\end{figure}  
\begin{figure}
\center{
\subfigure{\includegraphics[scale=0.6]{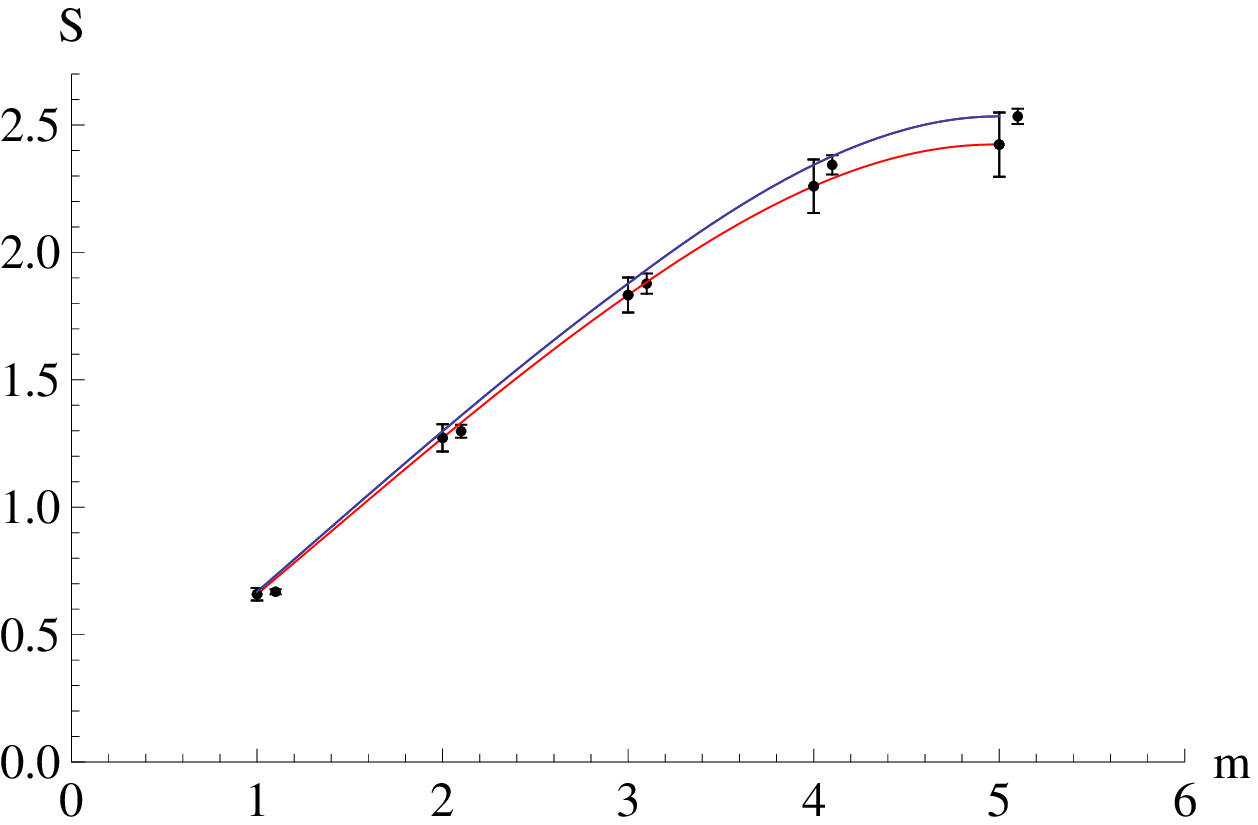}}
\subfigure{\includegraphics[scale=0.6]{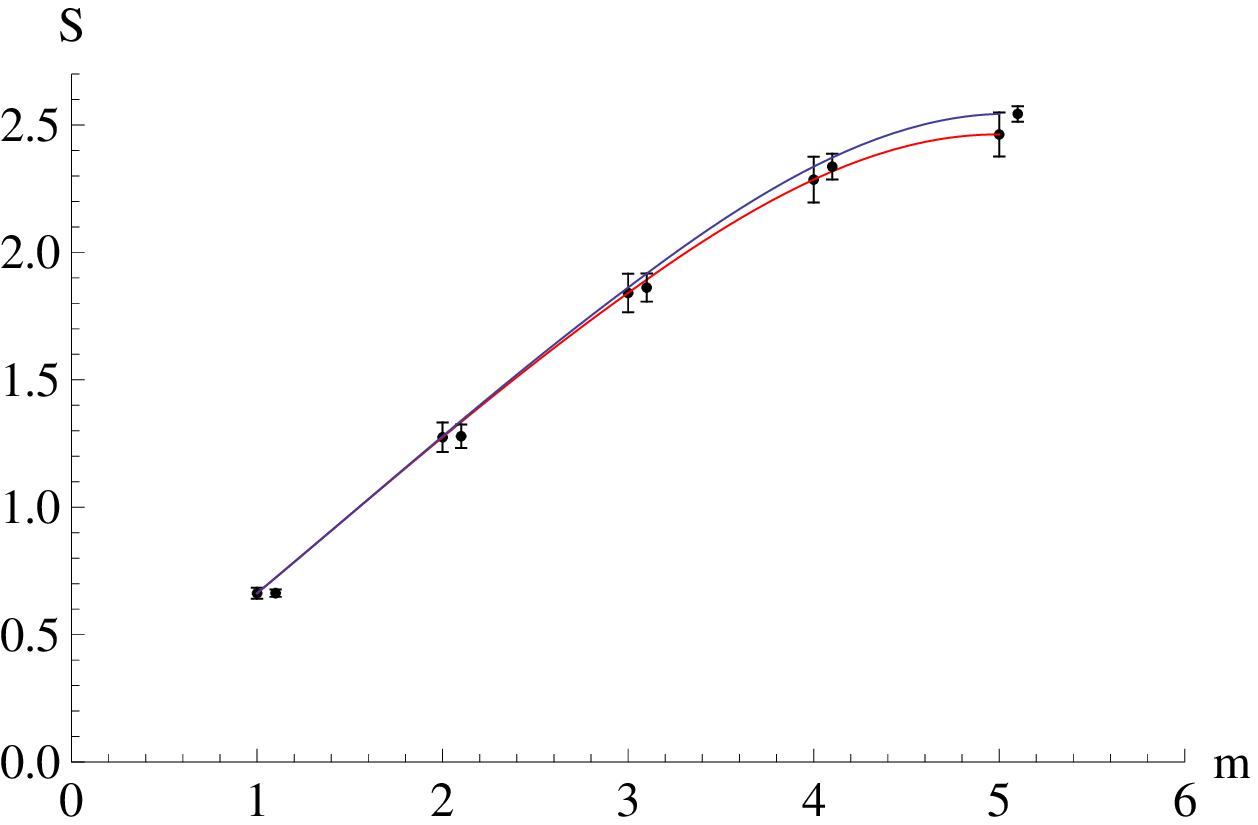}}
}
\caption{The subsystem size $m$ dependence of the bipartite entanglement, which is measured by von Neumann entropy for the superposition states $|\Psi(t)\rangle$ at $t=0,1,2,...,99$ (blue) and energy eigenstates $|E_n\rangle$ with $151\leq n\leq 200$ (red). The parameters are the same as those of Fig. 2, except for $\gamma_i$. (a)The case of uniform $z$-component of the magnetic field $\gamma_i=0.5$. (b)The $z$-component of magnetic field is randomly distributed in $[0,\Delta]$ with $\Delta=0.5$. We have confirmed that the entanglement entropy is quantitatively similar also for $\Delta=1$. The error bars for the superposition states are slightly shifted in the horizontal direction, and curves are just for the eye.}
\end{figure}
\begin{figure}
\center{
\subfigure[]{\includegraphics[scale=0.6]{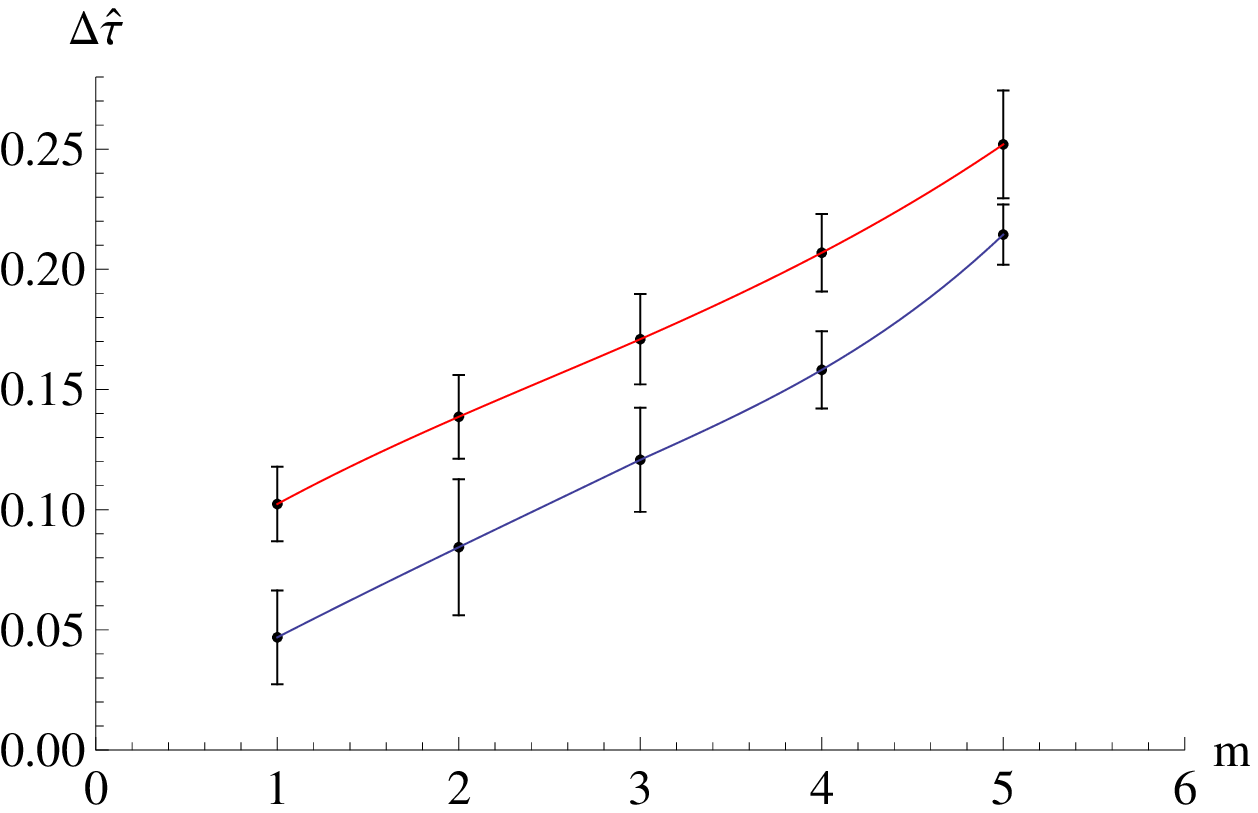}}
\subfigure[]{\includegraphics[scale=0.6]{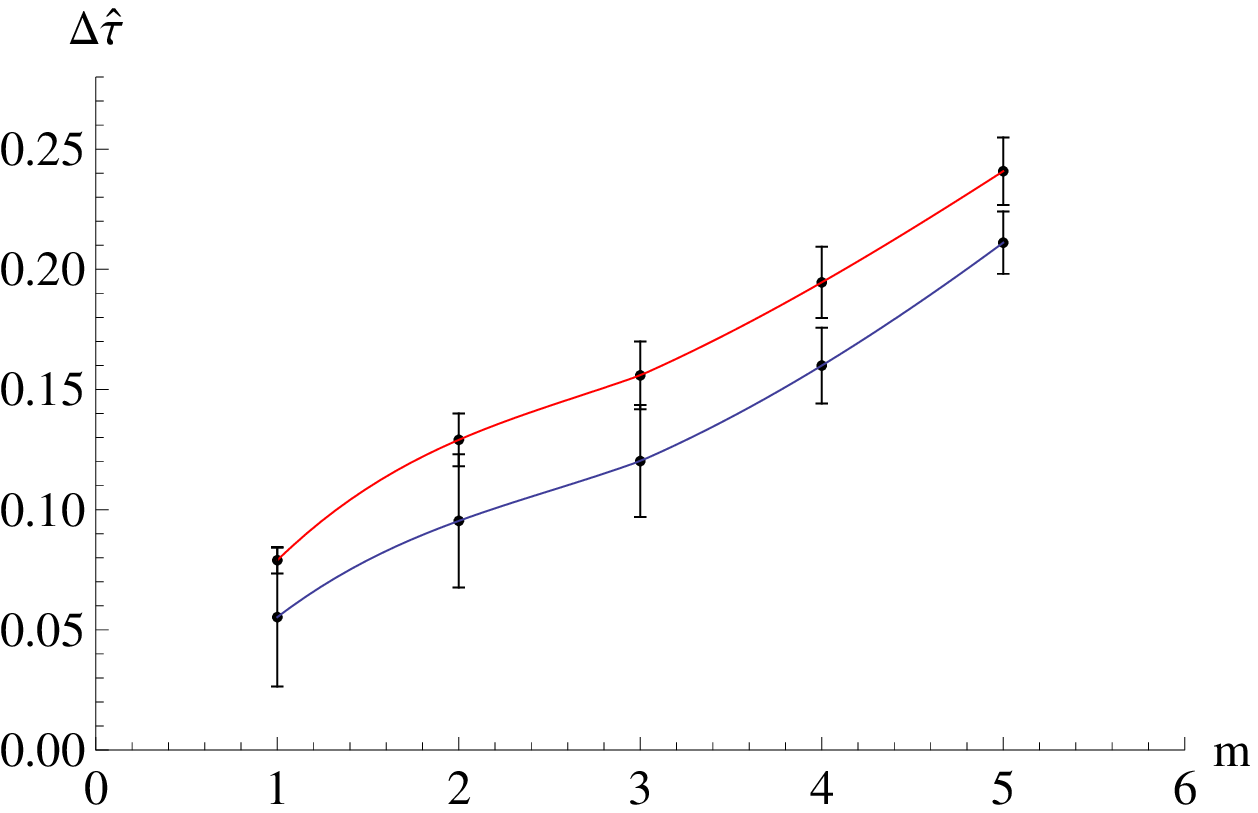}}
\subfigure[]{\includegraphics[scale=0.6]{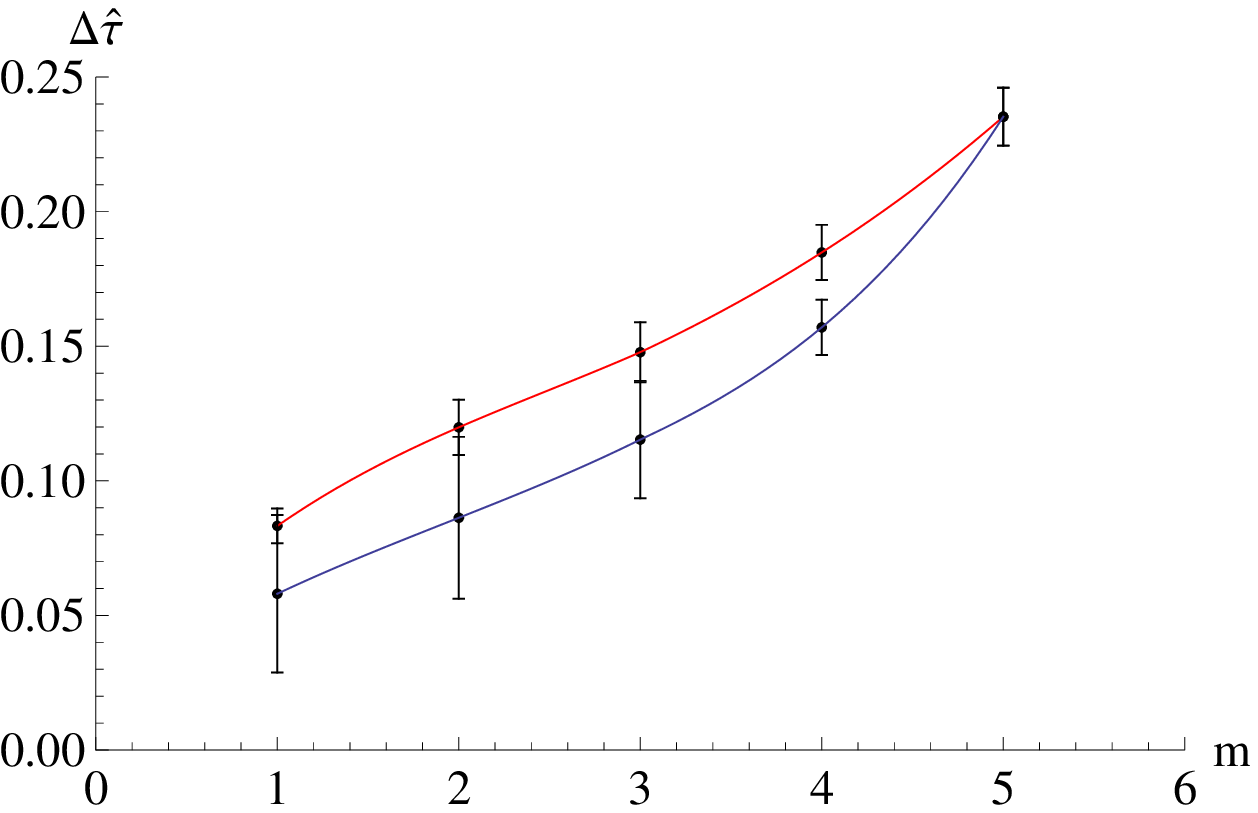}}
\subfigure[]{\includegraphics[scale=0.6]{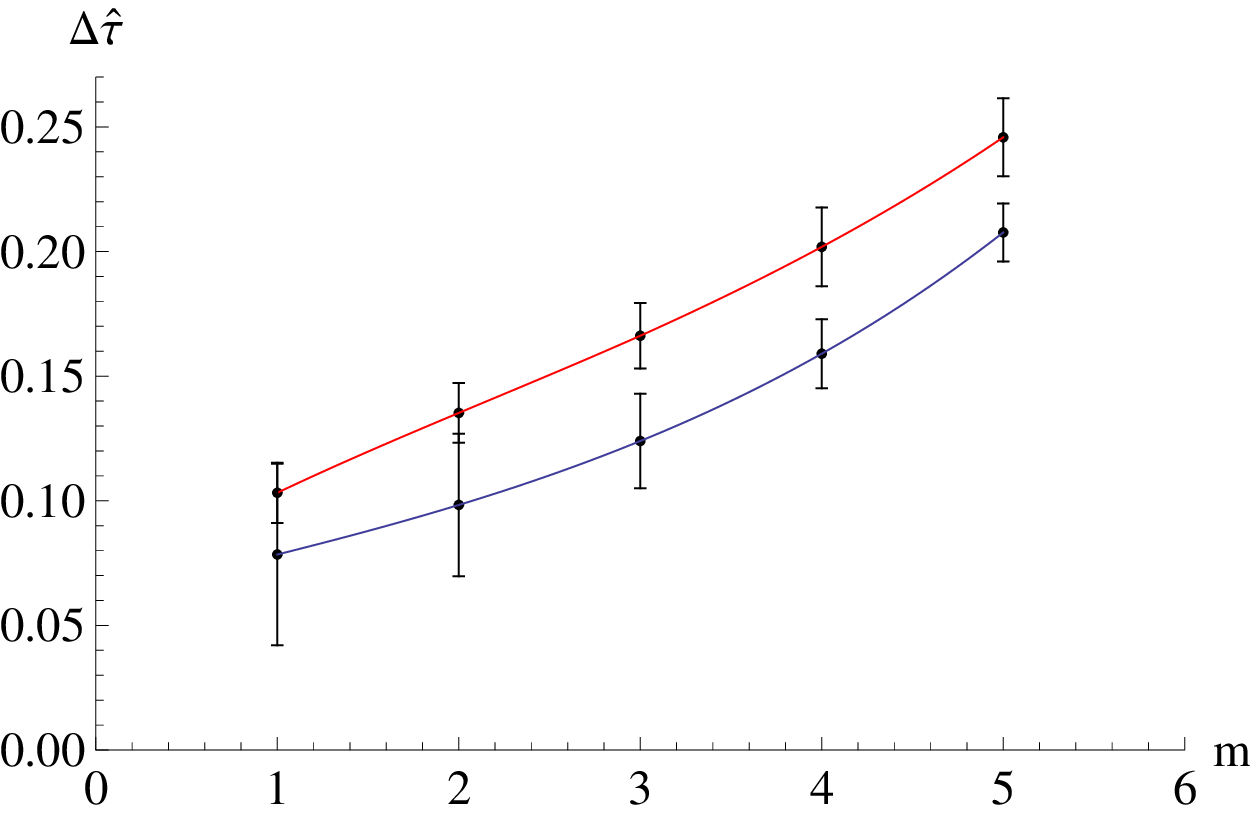}}
}
\caption{Subsystem size $m$ dependence of the deviations from averaged state $\Delta\hat{\rho}$ and $\Delta\hat{\tau}$ for the reduced density matrices of quasi eigenstates and energy eigenstates. (a) The case of uniform magnetic field in $z$ direction $\gamma_i=0.5$. We also explored random $\gamma_i$, which is  sampled from (b)$[0,0.5]$ and (c)$[0,1]$. (d) The case of XY model with $J=1$, $\Delta J=0.2$, and $\alpha=\gamma=0.5$.}
\end{figure}    
We also investigated the XY spin-chain model, whose Hamiltonian is $\hat{H}=-\sum_{j=1}^{N-1}\left((J+\Delta J)\hat{\sigma}_j^x\hat{\sigma}_{j+1}^x+(J-\Delta J)\hat{\sigma}_j^y\hat{\sigma}_{j+1}^y\right)+\alpha\sum_{j=1}^N\hat{\sigma}_j^x+\gamma\sum_{j=1}^N\hat{\sigma}_j^z$
 and confirmed that the subsystem size dependences of typical states $\Delta\hat{\rho}$ and eigenenergy states $\Delta\hat{\tau}$ are similar. For example, we show the case of $J=1$, $\Delta J=0.2$, and $\alpha=\gamma=0.5$ in Fig. 3(d).
For  
the integrable case $\gamma=0$\cite{Shankar1}, we calculated the distance $\Delta \hat{\tau}_n$ between $\tau_n$ and the averaged state $\hat{\rho}_0$ for $1\leq m\leq 5$. The distance is larger than the case of $\gamma=0.5$ roughly by a factor $1.5$ both for $N=9$ and $N=10$. 
\section{Summary}
We have shown that each energy eigenstate can be seen as a typical state in the basis of quasi eigenstates $|\Psi(t)\rangle$ of ``time operator" $\hat{T}$. From operational point of view, we can consider the measurement of ``time operator" as an estimation of unknown parameter $t$ of a given $|\Psi(t)\rangle$. We remark that it is possible to formally define the ``phase operator" $\hat{\theta}$approximately canonical conjugate to the number operator $\hat{N}$ in a similar way by using the gauge transformation instead of the unitary evolution: Given a state $e^{i\theta\hat{N}}|\Psi\rangle$ with unknown $\theta$, we can estimate $\theta$ by measuring  so-obtained phase operator. 

On the other hand, the subtlety of the non-existence of $\hat{T}$ as an observable rigorously conjugate to the Hamiltonian amounts to the approximate orthogonality of quasi eigenstates: There is a minimum time resolution $\tau$ given by the Boltzmann time\cite{Monnai1,Goldstein1,Reimann2} both for integrable and nonintegrable systems. 
Our point is that the quasi eigenstate of ``time operator" is time evolved state $|\Psi(t)\rangle$, and thus for thermodynamic systems where equilibration occur, $|\Psi(t)\rangle$ is considered to be in equilibrium for most $t$ according to the typicality\cite{Sugita1,Reimann1,Lebowitz1,Popescu1} so that its typical superposition is expected to be also thermal. 
This fact strongly suggests that ETH for diagonal elements holds as long as most of the time evolved states $|\Psi(t)\rangle$ well reproduces microcanonical expectation values for a class of observables. 
We numerically verified this argument in two different ways by comparisons of  bipartite entanglements of superposition states $|\Psi(t)\rangle$ and energy eigenstates $|E_n\rangle$, and of the averaged errors $\Delta\hat{\rho}$ and $\Delta\hat{\tau}$ for nonintegrable systems. The entanglement entropies of energy eigenstates are almost the same as those of superposition states.
On the other hand, the agreement of averaged errors of the reduced states indicates that energy eigenstates and superposition states yield similar expectation values for the observables of the subsystem.   

Though our argument in this paper focuses on the diagonal elements, we briefly mention on the off-diagonal elements, which are often supposed typically of order ${\cal O}(\frac{1}{d})$ in a version of ETH\cite{Anza2}. It remains as unsolved problem to explain the evaluation of off-diagonal elements in terms of ``time operator", and we explain another approach. The off-diagonal elements of observables $|\langle E_n|\hat{A}|E_m\rangle|^2$ are evaluated as order ${\cal O}(\frac{1}{d})$ by expanding $|E_n\rangle=\sum_{k=1}^dc_k|\phi_k\rangle$ and $|E_k\rangle=\sum_{k=1}^d d_k|\phi_k\rangle$ with a fixed basis $|\phi_k\rangle$ are two independent and mutually orthogonal typical states in the $d$-dimensional energy shell, i.e. we regard the coefficients $c_k$ and $d_k$ as random variables with respect to the Haar measure. Then, the average of the off-diagonal elements are considered to be ${\cal O}(\frac{\|\hat{A}\|^2}{d})$, which reproduces the ETH also for off-diagonal elements. 

In the presence of strong spatial disorder, ETH breaks down\cite{Huse1,Fan1,Hosur1}.
To explore the case of non-thermal case including the many-body localization possibly in more than one dimensions is an important future problem. 
\ack
\section{Acknowldgement}
The author is grateful to Professor K. Yuasa for fruitful discussions. 
This work is supported by Grants-in-Aid for Scientific Research (C) (No.\ 18K03467) from the Japan Society for the Promotion of Science (JSPS).
\\

\end{document}